# Heterogeneity of Synaptic Input Connectivity Regulates Spike-based Neuronal Avalanches

Shengdun Wu[a, #], Yangsong Zhang[a, b,#], Yan Cui[a], Heng Li[a], Jiakang Wang[a], Lijun Guo[a], Yang Xia[a, b], Dezhong Yao[a, b,*], Peng Xu[a, b], and Daqing Guo[a, b,*]

[a]The Clinical Hospital of Chengdu Brain Science Institute, MOE Key Lab for Neuroinformation, University of Electronic Science and Technology of China, Chengdu 611731, Peoples Republic of China

[b]School of Life Science and Technology, Center for Information in Medicine, University of Electronic Science and Technology of China, Chengdu 611731, Peoples Republic of China

## Abstract

Our mysterious brain is believed to operate near a non-equilibrium point and generate critical self-organized avalanches in neuronal activity. Recent experimental evidence has revealed significant heterogeneity in both synaptic input and output connectivity, but whether the structural heterogeneity participates in the regulation of neuronal avalanches remains poorly understood. By computational modelling, we predict that different types of structural heterogeneity contribute distinct effects on avalanche neurodynamics. In particular, neuronal avalanches can be triggered at an intermediate level of input heterogeneity, but heterogeneous output connectivity cannot evoke avalanche dynamics. In the criticality region, the co-emergence of multi-scale cortical activities is observed, and both the avalanche dynamics and neuronal oscillations are modulated by the input heterogeneity. Remarkably, we show similar results can be reproduced in networks with various types of in- and out-degree distributions. Overall, these findings not only provide details on the underlying circuitry mechanisms of nonrandom synaptic connectivity in the regulation of neuronal avalanches, but also inspire testable hypotheses for future experimental studies.

\# These authors contributed to this work equally

* Corresponding author: D.G. (email: dqguo@uestc.edu.cn)

# 1. Introduction

Cortical neurons continually integrate massive amounts of excitatory and inhibitory inputs from presynaptic neurons, and then produce diverse spatiotemporal patterns of collective activity (Brunel & Wang, 2003; Buzsaki, 2006). As a vital type of spatiotemporal pattern generated in the brain, neuronal avalanches have been widely observed in many experimental and computational studies (Friedman, et al., 2012; Plenz & Thiagarajan, 2007; Yu, et al., 2011). Similar to avalanches that emerge in other dynamical systems, the spatial and temporal distributions of neuronal avalanches have been identified as following power-law statistics, implying that this brain state operates near a non-equilibrium critical point (Bak, et al., 1987; Chialvo, 2010; Friedman, et al., 2012). Importantly, the criticality of neuronal ongoing activity has been seen across different brain spatial scales in experimental data. Using *in vitro* and *in vivo* recordings, avalanches of neuronal activity have been discovered at relatively small spatial scales at both spike and local field potential (LFP) levels (Beggs & Plenz, 2003, 2004; Bellay, et al., 2015; Ribeiro, et al., 2010; Shew, et al., 2009). Furthermore, previous studies have also shown the whole-brain activity dynamics measured with noninvasive techniques, such as electroencephalography (EEG) and functional magnetic resonance imaging (fMRI), can also be well described by power-law statistics (Linkenkaer-Hansen, et al., 2001; Stam & De Bruin, 2004). More intriguingly, neuronal avalanches have been postulated to facilitate information processing in the brain. For instance, it has been reported that neuronal avalanches can optimize the neuronal dynamic range, maximize the amount of information that can be transmitted and stored, and improve sensitivity to sensory inputs (Beggs & Plenz, 2003; Haldeman & Beggs, 2005; Kinouchi & Copelli, 2006; Shew & Plenz, 2013; Shew, et al., 2009; Shew, et al., 2011).

Understanding the dynamical and emergent mechanisms of neuronal avalanches may provide deep insights into the computational principles in the brain. Previous animal studies both *in vitro* and *in vivo*, together with computational modeling, have strongly suggested that the avalanche dynamics in neural systems may arise at the critical state in excitation-inhibition balanced networks and can be regulated by several intrinsic network properties, such as short-term synaptic plasticity and the balance level between excitation and inhibition (Beggs & Plenz, 2003; Levina, et al., 2007; Lombardi, et al., 2012; Lubenov & Siapas, 2008; Rubinov, et al., 2011;

Shew, et al., 2009). Consistent with experimental observations, recent investigations have shown that avalanches of neuronal activity preferentially emerge at a moderately synchronized state of collective firing activity and might coexist with both irregular firing and stochastic oscillations (Gireesh & Plenz, 2008; Lubenov & Siapas, 2008; Yang, et al., 2017). Notably, this finding is of particular interest because the co-emergence of these multi-scale cortical activities has been believed to ensure the cost-efficient information capacity of the brain, further emphasizing the functional significance of avalanche dynamics in neuronal information processing (Yang, et al., 2017).

Recent statistical analysis of multi-neuron population recordings has revealed that neighboring neurons differ highly in synaptic connectivity, implying the existence of structural heterogeneity in wiring diagrams among neurons (Bonifazi, et al., 2009; Okun, et al., 2015; Shimono & Beggs, 2014; Song, et al., 2005). This class of neuronal heterogeneity is ubiquitous in the brain and can be observed both between and within cell types (Okun, et al., 2015). Theoretically, the variability in the synaptic connectivity of neurons leads to stochasticity at the population level, which may further affect the spatiotemporal patterns of collective firing activity. Such stochastic effects indicate that structural heterogeneity may be a potential factor in the regulation of neuronal avalanches. Although simulations have identified the emergence of avalanche dynamics in heterogeneous neuronal networks in different topologies, these studies have generally considered a simplified assumption of uniform heterogeneity in the connectivity of synaptic inputs and outputs (Larremore, et al., 2011; Pajevic & Plenz, 2009). However, the assumption of uniform heterogeneity seems to be questionable because biological neurons may exhibit different degrees of structural heterogeneity in their input and output connectivity, an idea evidenced by accumulating observations in geometric and functional organization of cortical circuits (Chen, et al., 2013; Okun, et al., 2015). It remains controversial whether heterogeneous input and output connectivity contribute equally to evoke and regulate the critical dynamics in the brain.

To address this issue, we established a biologically plausible neural circuit model to evaluate the precise effects of heterogeneous input and output connectivity on the regulation of neuronal avalanches. Through computational modeling, we show that heterogeneous input connectivity can evoke and regulate spike-based neuronal avalanches at an appropriate level of heterogeneity by tuning the dynamics of neuronal ensembles, whereas heterogeneity in output connectivity influences the network dynamics slightly and cannot trigger neuronal avalanches. Importantly, we reproduce the similar phenomena by introducing different types of statistical distributions into the heterogeneous connections of the model, further demonstrating the generalizability of our results. These findings thus highlight the functional role of input heterogeneity in evoking and regulating critical dynamics in the brain.

## 2. Model and methods

### 2.1. Neural circuit model with heterogeneous connections

We attempt to propose generalized dynamical and regulation mechanisms for neuronal avalanches based on a neural circuit model with special heterogeneous structures. To this end, we establish a neuronal network with either heterogeneous input connectivity or heterogeneous output connectivity. Briefly, the network comprises a total of $N = 2000$ excitatory and inhibitory neurons with spiking dynamics. As schematically shown in Fig 1A, our model has a recurrent nature, which can be used to simulate the population dynamics of neurons from a local brain region. Similar to the excitatory-inhibitory ratio roughly observed in mammalian neocortex (Gerstner & Kistler, 2002), we employ $N_E = 1600$ excitatory neurons and $N_I = 400$ inhibitory neurons for all simulations.

Our strategy is to consider different types of structural heterogeneity in the topology of recurrent neuronal networks independently. To generate a specific network structure with heterogeneous input connectivity, the total number of presynaptic neurons to a target postsynaptic neuron, i.e., the in-degree, is assumed to be statistically distributed around its mean $K_{in}$ with a standard deviation $S_{in}$ (Fig 1B, top). Given the in-degree of the $i$-th neuron denoted by $K_{in}^i$, we assign it $K_{in}^i$ presynaptic partners randomly selected from other neurons in the network. In our model, we consider the link from the $i$-th neuron to the $j$-th neuron and the link from the $j$-th neuron to the $i$-th neuron as different synaptic connections. In addition, we also prohibit the existence of self-connections, and do not allow a neuron to be coupled to another neuron more than once. Mathematically, the mean and standard deviation of the in-degrees over all neurons

can be calculated as:

$$K_{\text{in}} = \frac{1}{N} \sum_{i=1}^{N} K_{\text{in}}^{i} \tag{1}$$

and

$$S_{\text{in}} = \sqrt{\frac{1}{N} \sum_{i=1}^{N} \left( K_{\text{in}}^{i} - K_{\text{in}} \right)^2}. \tag{2}$$

At the network level, the level of structural heterogeneity is determined by both the type of in-degree distribution and the parameter values of $K_{\text{in}}$ and $S_{\text{in}}$ (Fig 1B, top). For a given network density (i.e., $\rho = K_{\text{in}}/N$), a relatively larger $S_{\text{in}}$ indicates a wider in-degree distribution. Wider distributions correspond to higher levels of structural heterogeneity in the established neuronal network (Fig 1C). Compared to heterogeneous input connectivity, the level of output heterogeneity in such a network is relatively low because the presynaptic neurons are randomly assigned in our model. It is obvious that a similar constructing algorithm can be used to generate a network dominated by heterogeneous output connectivity as well, with the mean and standard deviation of the out-degrees represented by $K_{\text{out}}$ and $S_{\text{out}}$ (Fig 1B, bottom).

In simulations, we consider three kinds of statistical distributions for networks with heterogeneous input and output connectivity, including exponential, Gaussian and uniform distributions. In modeling studies, these types of statistical distributions have been widely used to construct neuronal networks. For simplicity, we perform experiments and summarize the main findings based on shifted exponential in- and out-degree distributions, and then further test our results using the Gaussian and uniform degree distributions. Considering the finite network size, we here truncate the degree distributions to make sure that the range of degrees is between 1 and 1999. Owing to the truncated effect, a certain error of the mean and standard deviation of degree distributions is unavoidable when the value of $S_{in}$ is very large, but it is still acceptable. Unless otherwise noted, we use the default parameter values listed in Table 1 to establish the neuronal network.

## 2.2. Single-neuron dynamics and simulation protocol

The spiking dynamics of neurons is mimicked using the leaky integrate-and-fire (LIF) model neuron with conductance-based synaptic current. The subthreshold membrane potential of each LIF neuron can be described as:

$$C_{\mathrm{m}} \frac{dV_i}{dt} = -G_{\mathrm{L}}(V_i - V_{\mathrm{L}}) + I_{\mathrm{syn}}^i(t) + I_{\mathrm{back}}^i(t). \tag{3}$$

Here $C_{\mathrm{m}}$ is the membrane capacitance, $V_i$ represents the membrane potential of the $i$-th neuron, $G_{\mathrm{L}}$ is the membrane leak conductance, and $V_{\mathrm{L}}$ is resting potential, and $I_{\mathrm{syn}}^i(t)$ denotes the total synaptic current. In addition, each neuron in the network is also driven by an independent background current to maintain network activity. In this study, the independent background current is mimicked as: $I_{\mathrm{back}}^i(t) = I_0 + \sigma \xi_i(t)$, where $I_0$ is the bias current, $\xi_i(t)$ is the Gaussian white noise with a zero mean and unit variance (here the unit of $\xi_i(t)$ is $\mathrm{nA} \bullet \mathrm{ms}^{1/2}$), and $\sigma$ is a dimensionless parameter representing the fluctuation intensity of the background current. When the membrane potential of a neuron reaches the spike threshold $V_{\mathrm{th}}$ from below, an action potential is emitted, and then the membrane potential is clamped at the resting potential $V_{\mathrm{rest}}$ for a short refractory period $\tau_{\mathrm{ref}}$.

For each neuron, we assume that the total synaptic current is conductance-based, modeled by:

$$I_{\mathrm{syn}}^i(t) = \sum_j G_{\mathrm{E}}(i,j)(V_i - V_{\mathrm{E}}) + \sum_k G_{\mathrm{I}}(i,k)(V_i - V_{\mathrm{I}}). \tag{4}$$

In this equation, the first and second outer sums include all excitatory and inhibitory synapses onto the $i$-th neuron, $G_{\mathrm{E}}(i,j)$ is the excitatory synaptic conductance from the $j$-th neuron to the $i$-th neuron, $G_{\mathrm{I}}(i,k)$ is the inhibitory synaptic conductance from the $k$-th neuron to the $i$-th neuron, and $V_{\mathrm{E}}$ and $V_{\mathrm{I}}$ are the reversal potentials for excitatory and inhibitory synapses, respectively. To simulate synaptic transmission, let us consider the $h$-th neuron as a presynaptic neuron of the $i$-th neuron. When this neuron emits an action potential, the corresponding postsynaptic conductance is increased after a transmission delay $\tau_{\mathrm{d}}$: $G_{\mathrm{E}}(i,h) \leftarrow G_{\mathrm{E}}(i,h) + \Delta G_{\mathrm{E}}$ for excitatory coupling, and $G_{\mathrm{I}}(i,h) \leftarrow G_{\mathrm{I}}(i,h) + \Delta G_{\mathrm{I}}$ for inhibitory coupling. Otherwise, the excitatory and inhibitory synaptic conductances decay exponentially:

$$\frac{d}{dt} G_{\mathrm{E}}(i,h) = -\frac{1}{\tau_{\mathrm{E}}} G_{\mathrm{E}}(i,h) \tag{5}$$

and

$$\frac{d}{dt} G_{\mathrm{I}}(i,h) = -\frac{1}{\tau_{\mathrm{I}}} G_{\mathrm{I}}(i,h) \tag{6}$$

where $\tau_E$ and $\tau_I$ are excitatory and inhibitory synaptic time constants, respectively. The coupling strengths $\Delta G_E$ and $\Delta G_I$ represent the relative peak conductances of excitatory and inhibitory synapses, and their values are determined by the types of pre and postsynaptic neurons. In simulations, we set $\Delta G_E = W_{EE}$ and $\Delta G_I = W_{EI}$ if the $i$-th neuron is an excitatory postsynaptic neuron, and choose $\Delta G_E = W_{IE}$ and $\Delta G_I = W_{II}$ when the neuron is an inhibitory postsynaptic neuron.

The stochastic differential system is solved by using the Euler-Maruyama method, with a relatively small temporal resolution of 0.1 ms to ensure an accurate simulation (Kloeden, et al., 2012). The default parameter values of the LIF neuron and synaptic model are listed in Table 1.

## 2.3. Data analysis

Each simulation is performed for a long time (up to 200 seconds) to record sufficiently data for further statistical analysis. Several data analysis methods are employed to quantitatively evaluate the spike data generated by our model. To calculate some neuronal measurements, we carry out 50 trials of simulations with different global random seeds. Considering that the population activity of inhibitory neurons exhibits a similar trend as that of excitatory neurons, we analyze only spike data recorded from excitatory neurons (see Fig 2, for example). Similar qualitative results also can be observed using inhibitory spike data.

**Analysis of network dynamics.** To assess the temporal regularity of spike trains at the network level, we compute the coefficient of variation of inter-spike intervals ($CV_{ISI}$) over all excitatory neurons. Mathematically, the $CV_{ISI}$ for the $i$–th neuron is defined as (Holt, et al., 1996):

$$CV_{ISI}^i = \frac{\sqrt{\langle T_j^2 \rangle - \langle T_j \rangle^2}}{\langle T_j \rangle}. \tag{7}$$

Here the symbol $\langle \bullet \rangle$ represents the average over time, $T_j = t_{j+1} - t_j$, and $t_j$ is the time of $j$-th spike. Then, the mean $CV_{ISI}$ of all excitatory neurons is given as:

$$CV_{ISI} = \frac{1}{N_E} \sum_{i=1}^{N_E} CV_{ISI}^i. \tag{8}$$

By definition, a smaller value of $CV_{ISI}$ corresponds to a relatively better temporal regularity of spike trains at the network level.

The synchronization of population activity is estimated by a coherence index (X.-J. Wang, 2002). To compute this measurement, the instantaneous population firing rate of excitatory neurons $R(t)$ is calculated in 2.0-ms bins. Then, we can obtain both the mean and standard deviation of $R(t)$ over time, which are denoted by $\mu_R$ and $\sigma_R$, respectively. As a dimensionless measurement, the coherence index is finally defined as:

$$\text{Syn} = \frac{\sigma_R}{\mu_R}. \tag{9}$$

Obviously, the larger the value of $\text{Syn}$, the better the synchronization in the network.

In this work, neuronal oscillations are simply represented by the collective firing of excitatory neurons computed from 0.5-ms bins (the sampling rate: $f_s = 2000$ Hz) and measured with spectral analysis. To visualize the typical features of neuronal oscillations in both the time and frequency domains, we use a Complex Morlet wavelet transform to calculate the continuous power spectrograms for the average firing rate of excitatory neurons. The default values of the bandwidth parameter and wavelet center frequency in such time-frequency spectral analysis are fixed at 1.5 and 0.8 Hz, respectively. Moreover, we employ power spectral analysis to evaluate the neuronal oscillations generated by the model. To do this, the power spectral density of the average firing rate of excitatory neurons is obtained by using the Welch's method with a 2048-sample Hamming widow and 50% overlap. Then, both the peak power level of the neuronal oscillations and the corresponding peak frequency can be captured from the curve of power spectral density.

**Analysis of spike-based neuronal avalanches.** For analyzing neuronal avalanches, we sample the spike data from $N_S$ randomly selected excitatory neurons and bin the population activity into time windows ($\Delta t = 0.5$ ms). This sampling process is repeated 20 times for each simulation, and the overall data are used for further statistical analysis. An underlying avalanche event is defined as a sequence of time bins in which at least one spike is emitted, ending with a silent time bin. For a given avalanche event, we denote its size $S$ and duration $T$ as the total number of spikes contained in this event and the corresponding lifetime of this event. By using all avalanche events recorded in a specific experiment, we can estimate the probability distributions of avalanche size and duration, represented by $P(S)$ and $P(T)$, respectively.

To characterize neuronal avalanches, we calculate the distances from the real distributions $P(S)$ and $P(T)$ to their corresponding best-fitted power-law distributions $P_{\text{fit}}(S) \sim S^\alpha$ and $P_{\text{fit}}(T) \sim T^\beta$. Mathematically, these two distance measurements are given as (Yang, et al., 2017):

$$D_S = \frac{\sum_S S |P(S) - P_{\text{fit}}(S)|}{\left|\sum_S S |P_{\text{fit}}(S)|\right|} \tag{10}$$

and

$$D_T = \frac{\sum_T T |P(T) - P_{\text{fit}}(T)|}{\left|\sum_T T |P_{\text{fit}}(T)|\right|}. \tag{11}$$

For both avalanche size and duration, we numerically solve their best-fitted power-law distributions based on the lowest distance values. Obviously, the optimal slopes $\alpha$ and $\beta$ are also determined at the lowest distance values. The lower the distance value, the closer the probability distribution of the real data is to the power-law distribution. In the present study, we consider that our model operates in the criticality region when $D_S < 0.4$. Note that a similar criticality region can also be observed for $D_T < 0.35$. In some cases, we also compute the average size $\langle S \rangle (T)$ conditioned on a given duration $T$. With the least square method, the optimal slope $\pi$ can be also estimated by best-fitting the power-law distribution $\langle S \rangle (T) \sim T^\pi$. Based on the universal scaling theory, it can be predicted that the model operates near a non-equilibrium point when three critical exponents satisfy the expected relation: $\pi = (1 + \beta)/(1 + \alpha)$.

# 3. Results

## 3.1. Impacts of different types of structural heterogeneity on neurodynamics

As a preliminary step, we ask whether heterogeneous input and out connectivity contribute equally to the modulation of network dynamics. To answer this question, two representative simulations are performed by suddenly changing the levels of structural heterogeneity in networks with input and output connectivity. Surprisingly, we observe distinct dynamical behaviors emerging in networks with different types of structural heterogeneity (Fig 2). At a low level of structural heterogeneity, neurons in these two types of networks receive almost equal amounts of total synaptic currents with relatively strong intensities (Fig 2A and 2B). Under this condition, such strong synaptic interactions dominate the network dynamics and drive the collective activity of neurons to exhibit a rhythmic synchronous firing state. By introducing an intermediate level of heterogeneity into synaptic inputs, we find that the perfect synchronous firing of neurons is broken and a moderately synchronous firing state is evoked after a short transition period (Fig 2A, left shaded region). Theoretically, the emergence of moderate synchrony among neurons decreases the overall effect of their collective firing. Such a reduction not only destroys the temporal structures of synaptic currents, but also tends to enhance their stochastic fluctuations (Fig 2A). Nevertheless, a similar desynchronized behavior does not appear in networks with heterogeneous output connectivity (Fig. 2B). This finding suggests that output heterogeneity might not play a pivotal role in tempering neurodynamics. Further increasing the structural heterogeneity in these two types of networks demonstrates that the network dynamics are heavily influenced only by the input heterogeneity (Fig 2A and 2B). With a strong level of input heterogeneity, we observe that the network dynamics may be even pushed into an asynchronous irregular firing state in the model (Fig. 2A).

To understand why heterogeneous input and output connectivity show differential contributions to the modulation of network dynamics, we compare several key neuronal measurements in these two types of networks (Fig 3). Figure 3A shows the average standard deviation of the total synaptic currents received by each neuron at different levels of structural heterogeneity. As expected, increasing the input heterogeneity in the network significantly enhances the variability in synaptic currents. In principle, a strong variability in synaptic currents introduces a high level of stochasticity into neurons, thus leading to a reduction in network synchronization and triggering intrinsic irregular neuronal firing. These dynamical behaviors correspond to the low coherence index seen in Fig 3B and the high mean coefficient of variation of inter-spike intervals ($CV_{ISI}$) observed in Fig 3C. Such desynchronized behavior might also suppress the effect of collective firing activity and, therefore, a decrease in average firing rate appears when there is a high level of input heterogeneity. Although strengthening the output heterogeneity also changes the network structure, it results in almost no additional variability in synaptic currents (Fig 3A). Thus, heterogeneous output connectivity only slightly impacts collective neuronal firing, which is quantitatively confirmed by other neuronal measurements, including the coherence index, the mean $CV_{ISI}$ and the average firing rate (Fig 3B-3D). As a consequence, the highly synchronous firing of neurons can be well preserved in our model even at sufficiently strong levels of output heterogeneity (Fig 2B).

These findings consistently reveal that not all types of structural heterogeneity contribute to the dynamics of neuronal ensembles. In particular, we identify that the heterogeneity of synaptic inputs is an important underlying factor that performs a functional role in regulating the collective firing of neuronal ensembles.

## 3.2. Heterogeneous input connectivity evokes and regulates neuronal avalanches

Recent experimental and computational studies have emphasized that avalanches of neuronal activity may emerge when there is moderately synchronous firing of neurons (Gireesh & Plenz, 2008; Lubenov & Siapas, 2008; Yang, et al., 2017). To examine whether the critical neuronal avalanches can also be triggered by the structural heterogeneity in synaptic connectivity, we perform spike-based analysis by calculating the size and duration of each avalanche event (Fig 4A, and see Model and methods) in the present study. For comparison with real electrophysiological experiments, we collect avalanche events from a small population of randomly chosen excitatory neurons (the sampling size $N_S = 400$ for default), but not all neurons, in the network.

Figure 4B shows the distributions of avalanche size (top) and avalanche duration (bottom) for networks with input heterogeneity. As we can see, the model exhibits distinct avalanche dynamics at different levels of structural heterogeneity. For an appropriate level of input heterogeneity, the distributions of both avalanche size and duration obey linear relationships in log-log coordinates (Fig 4B, black lines). These two linear relationships can be well characterized by the exponents of power-law statistics ($\alpha = -1.61$ and $\beta = -1.8$). Remarkably, the further plotting of the average avalanche size as a function of avalanche duration in the log-log coordinates reveals another power-law scaling, with an exponent of $\pi = 1.31$ (Fig 4B, inset). These exponents satisfy the expected relation $\pi = (1 + \beta)/(1 + \alpha)$ as predicted for a critical system by the scaling theory of non-equilibrium critical phenomena (Rybarsch & Bornholdt, 2014; Sethna, et al., 2001). In agreement with previous *in vivo* and *in vitro* experiments, these findings together suggest the occurrence of avalanche dynamics in our model (Beggs & Plenz, 2003, 2004;

Bellay, et al., 2015; Ribeiro, et al., 2010; Shew, et al., 2009). However, both increasing and decreasing the input heterogeneity can push the self-organized criticality of the system into other dynamical states. A significant reduction of input heterogeneity causes neurons to fire in a highly synchronous manner, which increases the chance of large-size avalanches and evokes supercritical avalanche dynamics (Fig. 4B, red lines). In contrast, the increase in input heterogeneity may lead to asynchronous irregular neuronal firing, triggering subcritical dynamics with exponential-type distributions of avalanche size and duration (Fig 4B, blue lines). However, a similar tuning effect is not observed in networks with heterogeneous output connectivity (Fig 4C). Theoretically, this finding is not surprising because variability in synaptic outputs has been shown to contribute weakly to the collective neuronal firing. The above results thus indicate that a suitable level of input heterogeneity can evoke neuronal avalanches in the brain.

To characterize the performance of avalanche dynamics, we compute distances between real and best-fitted power-law distributions for both avalanche size and duration at different levels of input heterogeneity (Fig 5A, and see Model and methods). Obviously, a smaller distance implies better network performance towards to the self-organized criticality. As the input heterogeneity is increased, both distance measurements first drop and then rise, and neuronal avalanches occur at intermediate levels of input heterogeneity (Fig 5A, shaded region). This evidence further demonstrates that appropriately heterogeneous input connectivity can trigger critical dynamics in neural systems. Moreover, fitting the power-law statistics for avalanche events also reveals that the slope parameter $\alpha$ (avalanche size) is dramatically impacted by $S_{in}$, but the slope parameter $\beta$ (avalanche duration) shows insensitivity to the input heterogeneity (Fig 5B). In the criticality region, both of these values are close to -1.5, and the slope $\alpha$ displays decreasing values with increased input heterogeneity (Fig 5B, shaded region). By assessing the relationships between the two distance measurements and the coherence index, we find that neuronal avalanches appear at moderately synchronous levels of neuronal firing (Fig 5C), indicating the co-emergence of multi-scale activities generated by our model. Note that this observation might be biological significant, because the co-emerged cortical activities are believed to be essential for efficient information processing in the brain. Increasing the input heterogeneity not only decreases the strength of neuronal oscillations due to desynchronization, but also increases the peak oscillation frequency (Fig 5D and 5E). Interestingly, such modulation of neuronal oscillations occurs at the low beta band (13-20 Hz). We highlight these results because neuronal oscillations in the beta band have been widely observed in experimental recordings and are believed to play important roles in both cognitive processing and working memory (Engel & Fries, 2010; Gireesh & Plenz, 2008).

Our model confirms that the heterogeneity in synaptic inputs can not only evoke spike-based neuronal avalanches and induce the co-emergence of multi-scale cortical activities, but also regulate the intrinsic properties of neuronal avalanches.

### 3.3. Roles of sampling size in the neuronal avalanche observations

Previous experimental studies have shown that randomly removing events from both spike- and LFP-based recordings can significantly impact the distributions of the underlying neuronal dynamics, thus suggesting that subsampling might prevent the observation of characteristic power-law statistics (Hahn, et al., 2010; Petermann, et al., 2009; Priesemann, et al., 2009). However, due to limitations in current experimental techniques, neuronal data used in these studies are recorded from limited numbers of electrodes or neurons. It is still unknown whether the oversampling of neuronal data may also destroy the power-law distributions of neuronal avalanches.

To explore both the effects of subsampling and oversampling in our simulated data, we alter the number of the sampled excitatory neurons to control the sampling level in this study. Using the spike data generated in the same experiment (Fig. 4B, $S_{in} = 240$), we recalculate the distributions for both avalanche size and duration at different sampling sizes. As shown in Fig 6A and 6B, fine power-law statistics of avalanche events can be preserved for a relatively large range of intermediate sampling sizes. To a certain extent, this confirms the stability and consistency of the neuronal avalanches generated in our model. Further examinations of distance measurements versus sampling size reveal inverse bell-shaped profiles for both avalanche size and duration, indicating that either a too small or a sufficiently large number of samples tends to mask the power-law distributions of avalanche events (Fig. 6C). In partial agreement with experimental observations, recording insufficient spike-based events from a limited number of neurons might prevent the observation of the characteristic power-law due to the excessive loss of large-size avalanches (Fig 6A and 6B, $N_S = 200$). In contrast, collecting data from a very large population of neurons might both markedly reduce the chance of small-size avalanches and increase the possibility of large-size avalanches, resulting in a pseudo-supercritical behavior with gentle $\alpha$ and $\beta$ slopes (Fig. 6A and 6B, $N_S = 800$). This observation implies that the power-law statistics of neuronal avalanches might also be broken in cases of extreme oversampling. Indeed, similar shaping effects also exist for different intermediate sampling sizes. Consequently, we find that the best-fitted slopes for power-law distributions of avalanche size and duration are progressively increased towards to 0 with the growth of sampling size (Fig. 6D).

Our above findings demonstrate that both extreme subsampling and oversampling of neuronal data might prevent the observation of the power-law statistics of the underlying avalanche dynamics, and sampling size might serve as an important intrinsic parameter when identifying critical neurodynamics in experiments.

## 3.4. Our results can be extended to other in- and out-degree distributions

By assuming exponential in- and out-degree distributions, we have shown that the heterogeneity level in synaptic inputs significantly modulates network dynamics and can evoke spike-based neuronal avalanches under suitable conditions. A natural arising question is whether similar results can also be reproduced in networks with other types of degree distributions. To answer this question, we perform additional simulations considering both Gaussian and uniform distributions for heterogeneous input and output connectivity in our model. Similar to that of exponential distribution (Fig 2 and 3), our preliminary analysis confirms that heterogeneous input connectivity modulates the network dynamics in a significant way, whereas heterogeneity in output connectivity influences the network dynamics only slightly (data not shown). In Fig 7, we plot several typical distributions of avalanche events at different levels of input heterogeneity. For networks with both truncated Gaussian and uniform in-degree distributions, the spike-based avalanche events exhibit fine power-law distributions at intermediate levels of input heterogeneity, which can be characterized by three exponents satisfying the relation: $\pi = (1 + \beta)/(1 + \alpha)$. This theoretical signature indicates the occurrence of critical dynamics. Further examination of the distances between real and best-fitted power-law distributions versus the parameter $S_{in}$ reveals inverse bell-shaped profiles for both avalanche size and duration (Fig. 8A and 8B, left). As expected, our model displays critical avalanche dynamics only at an intermediate level of input heterogeneity for both Gaussian and uniform in-degree distributions (Fig. 8A and 8B, shaded regions). Compared with the case of uniform in-degree distribution, we find that networks with exponential and Gaussian in-degree distributions have relatively wider and right-shifted criticality regions. For a fixed network density, this difference might be because the true level of input heterogeneity is not only determined by the parameter $S_{in}$, but also impacted by the type of in-degree distribution itself.

Moreover, best-fitting the power-law distributions versus the parameter $S_{in}$ suggests that the exponent for avalanche size is modulated by the input heterogeneity as well (Fig 8A and 8B, middle). In the criticality region, the slope $\alpha$ decreases with increasing input heterogeneity for networks with Gaussian and uniform in-degree distributions. Replotting the distance measurements as a function of the coherence index shows that neuronal avalanches emerge in moderately synchronous firing states (Fig 8A and 8B, right). Obviously, this implies again that our model may generate the co-emergence of multi-scale activities in the criticality region. In addition, we observe that strengthening the input heterogeneity in our model can decrease the peak power of neuronal oscillations with a slightly enhanced dominant frequency (Fig 9, red lines). For networks with Gaussian and uniform out-degree distributions, the peak power of the neuronal oscillations exhibits a similar reducing trend with the growth of output heterogeneity, but the corresponding peak frequency shows insensitivity to output heterogeneity (Fig 9, blue lines).

Note that these findings are in agreement with our above observations for exponential in- and out-degree distributions. We thus postulate that the structural heterogeneity of input connectivity may be a generalized factor in the brain that plays a functional role in evoking and regulating neuronal avalanches.

## 4. Discussion

Avalanches of neuronal activity have been widely observed in electrophysiological recordings at different signal levels and brain spatial scales, but their dynamical mechanisms remain controversial (Beggs & Plenz, 2003; Bellay, et al., 2015; Ribeiro, et al., 2010; Yu, et al., 2011). Using a proof-of-principle neural circuit model that incorporates different types of in- and out-degree distributions, we proposed here a very simple, yet effective, mechanism to control the critical dynamics in local neural systems. In particular, we showed that heterogeneous input connectivity modulates the collective firing of neurons in a wide dynamic range (Landau, et al., 2016; Litwin-Kumar, et al., 2017), and that introducing an appropriate level of input heterogeneity into the model can suppress synchronous firing of neurons and trigger spike-based neuronal avalanches. Moreover, we demonstrated that tuning the level of input heterogeneity in the criticality region can not only vary the exponent of the power-law distributions for avalanche size, but also regulate the peak frequency and power of neuronal oscillations. These results highlight the functional importance of heterogeneity in synaptic inputs in evoking and tempering critical neurodynamics.

In contrast, we found that heterogeneous output connectivity might modulate collective neuronal firing only within a narrow dynamic range. In such a neuronal network with stubborn dynamics, it seems to be impossible to evoke neuronal avalanches by solely strengthening the output heterogeneity, possibly because our model operates at the oscillatory state, which is away from a bifurcation. In this case, tuning the level of output heterogeneity does not considerably affect the homogeneity of in-degrees in the network, and therefore will not introduce sufficient additional variability in synaptic currents to change the dynamical state. These results are in agreement with former computational observations, showing that broadening the out-degree distribution does not affect the dynamical state of a neuronal network operating away from a bifurcation (Roxin, 2011). However, if our model is poised near a non-equilibrium critical point, broadening the out-degree distribution might result in qualitative changes in the dynamical state of the network (Roxin, 2011). Under this condition, we infer that changing the level of output heterogeneity may also regulate the avalanches of neuronal activity in a relatively strong manner. This hypothesis deserves further examination in our further modelling studies.

Neural computations in the brain require fast and energy-efficient information processing capabilities. Our results confirmed that spike-based neuronal avalanches are likely to appear at a moderately synchronized firing state and can coexist with both irregular firing and stochastic oscillations. These observations might provide important biological implications, because the similar co-emergence of multi-scale cortical activities has been observed in both experimental and computational recordings (Gireesh & Plenz, 2008; S.-J. Wang, et al., 2016; Yang, et al., 2017; Yu, et al., 2011). Remarkably, recent theoretical analysis on these multi-scale cortical activities have shown that the simultaneous appearance of these cortical activities may provide a dynamical substrate for efficient neuronal information processing with high flexibility and capacity (S.-J. Wang, et al., 2016). In addition, previous studies using information-based measurements have also suggested that the neural system may exhibit the optimal signal processing capability when it works in the criticality region (Beggs & Plenz, 2003, 2004; Kinouchi & Copelli, 2006; Shew, et al., 2009; Shew, et al., 2011). We thus postulate that real neural circuits must maintain a certain level of input heterogeneity, providing a plausible underlying biological mechanism for the energy-efficient information processing capacity of the brain.

There are two testable predictions that emerge from our current results. First, previous studies have demonstrated that subsampling may prevent the observation of characteristic power-law statistics, but the effect of oversampling on neuronal avalanche distributions is still poorly understood (Hahn, et al., 2010; Petermann, et al., 2009; Priesemann, et al., 2009). By altering the number of sampled excitatory neurons, our oversampling analysis suggests that excessive oversampling may also mask the power-law distributions of avalanche events. In principle, this hypothesis can be examined in future experiments using high-density electrode arrays. Second, it has been reported that the exponents of the power-law distributions for both avalanche size and duration are close to -1.5, but their values might vary slightly with different experimental settings and recorded data (Beggs & Plenz, 2003; Gireesh & Plenz, 2008; Hahn, et al., 2010; Palva, et al., 2013; Yu, et al., 2014). Our current results predict that the observations of these two exponents might be modulated by both the sampling size and the level of input heterogeneity, a prediction that can be further tested experimentally.

The model we established in this study assumes an inhomogeneous connectivity structure. In fact, the inhomogeneous connectivity structure is an intrinsic property of the brain, an idea supported by sufficient experimental data. For instance, it has been well established that the wiring diagrams among neurons are highly nonrandom, and several significantly recurring nontrivial patterns of interconnections, termed "motifs", are contained in neural circuits (Song, et al., 2005). Recent studies have also uncovered that cortical neurons have remarkable heterogeneity in their input connectivity and input synaptic currents (Okun, et al., 2015; Xue, et al., 2014). Moreover, analyzing the functional organization of population neurons has shown that both the in- and out-degree of synaptic connectivity are broadly distributed with heavy tails (Bonifazi, et al., 2009; Ito, et al., 2014; Shimono & Beggs, 2014; Timme, et al., 2016). Together, these investigations imply that structural heterogeneity is ubiquitous in the brain and can coexist in both synaptic inputs and outputs, thus providing the structural basis for our observations. By further performing reproducibility analyses, we have demonstrated that our key findings can be observed in networks with different in- and out-degree distributions. These findings indicate that our presented results are independent of the specific type of inhomogeneous connectivity structure. Consequently, we propose that both the emergence and regulation of neuronal avalanches caused by heterogeneous input connectivity might be a generalized mechanism in the brain.

However, it is worth noting that a neuronal network with heterogeneous input connectivity allows neurons to exhibit significant heterogeneity in their total synaptic current, which might be a true underlying factor that evokes and regulates neuronal avalanches. An interesting question is whether the heterogeneity in the total synaptic current can also be induced by other biological mechanisms. There are several underlying biological mechanisms that can achieve this function, and one of them is discussed below. Previous experimental data have revealed that synaptic strengths recorded *in vitro* are not homogeneous but broadly distributed in a lognormal distribution (Song, et al., 2005). By introducing such lognormal distributed synaptic strengths into a neuronal network, remarkable heterogeneity in the total synaptic current can be easily induced even for a homogeneous connectivity structure. Therefore, the highly inhomogeneous distribution of synaptic strengths might be an alternative regulator of critical neurodynamics. In principle, these two underlying regulators might cooperate together in our brain and offer a stable circuitry basis in support of neuronal avalanches.

Though our model predicts that heterogeneous output connectivity does not play a pivotal role in tampering avalanche dynamics, we cannot rule out other underlying functional roles of the heterogeneity of synaptic outputs in the brain. Indeed, recent studies have confirmed that heterogeneous output connectivity may be critical for neural computations in feedforward networks, and neurons that compute the most information tend to receive inputs from high-degree neurons (Timme, et al., 2016). It has also been reported that broadening the out-degree may increase the amplitude of the cross-correlation of synaptic currents in recurrent networks (Roxin, 2011). In the real brain, different types of structural heterogeneity may perform complementary roles and provide a hybrid mechanism to ensure optimal neural computations. After a long duration of evolution, it is reasonable to suppose that our brain might have the ability to use this type of hybrid mechanism to achieve complicated functions.

Our model is parsimonious, designed to capture the fundamental biophysical mechanisms of neuronal avalanches contributed by structural heterogeneity. The limitations of this model and several possible extensions are discussed below. First, we did not introduce any plasticity mechanisms in our model. Indeed, synaptic strength can be largely mediated by both pre- and postsynaptic firing activities, with a temporal span ranging from milliseconds to several days (Gerstner & Kistler, 2002). In particular, previous studies based on homogeneous connectivity structures have shown that short-term plasticity may significantly influence self-organized criticality in neural systems (Levina, et al., 2007). It is necessary to further probe whether the avalanche dynamics generated in inhomogeneous networks can also be modulated by short-term plasticity. Second, we consider only the structural heterogeneity of synaptic inputs and outputs independently in our current model. However, as mentioned above, both heterogeneous input and output connectivity should coexist in cortical circuits (Chen, et al., 2013; Okun, et al., 2015). In future studies, we should explore the possible combined roles of these two types of structural heterogeneity in the regulation of neuronal avalanches. Finally, our current results are based on spike data generated in local neural circuits, but neuronal avalanches have also been observed at relatively large spatial scales (Beggs & Plenz, 2003, 2004; Hahn, et al., 2010). By integrating multimodal neuroimaging data, recent developed computational modeling techniques allow us to establish a large-scale brain model with the neural-field theory (Deco & Jirsa, 2012; Deco, et al., 2013). Using such a large-scale brain model, it is possible to further examine the roles of different types of structural heterogeneity in the regulation of neuronal avalanches even at the whole-brain level.

In conclusion, we have performed mechanistic studies to investigate the roles of different types of structural heterogeneity in the control of neuronal avalanches. Our results emphasize the functional significance of heterogeneous input connectivity in mediating neurodynamics, and provide the first computational evidence that avalanches of neuronal activity can be evoked and regulated by the level of input heterogeneity. We hope that these findings might not only deepen our current understanding on the biophysical mechanisms of neuronal avalanches, but also inspire testable hypotheses for future experimental studies.

# Acknowledgments

We sincerely thank Prof. Bharat Biswal for insightful discussions, and acknowledge Dr. Ke Chen and Mingming Chen for valuable comments on the manuscript. This work is supposed by the National Natural Science Foundation of China (Grant Nos. 31771149, 61527815, 81571770, 81330032), the Project of Science and Technology Department of Sichuan Province (Grant Nos. 2017HH0001 and 2018HH0003), and the 111 project (B12027).

# Competing financial interests:

# Table legend

**Table 1 Default values of model parameters used in numerical simulations.**

| Symbol | Description | Value |
|---|---|---|
| $N$ | Total number of neurons | 2000 |
| $N_E$ | Total number of excitatory neurons | 1600 |
| $N_I$ | Total number of inhibitory neurons | 400 |
| $N_S$ | Sampling size | 400 |
| $K_{in}$ | Mean of in-degrees | 400 |
| $K_{out}$ | Mean of out-degrees | 400 |
| $\rho$ | Network density | 0.2 |
| $S_{in}$ | Standard deviation of in-degrees | 0-380 |
| $S_{out}$ | Standard deviation of out-degrees | 0-380 |
| $W_{EE}$ | Synaptic strength of E→E coupling | 0.06 nS |
| $W_{IE}$ | Synaptic strength of E→I coupling | 0.03 nS |
| $W_{EI}$ | Synaptic strength of I→E coupling | 0.3 nS |
| $W_{II}$ | Synaptic strength of I→I coupling | 0.07 Ns |
| $C_m$ | Membrane capacitance | 0.5 nF (E), 0.2 nF (I) |
| $C_L$ | Leak conductance | 25 nS (E), 20 nS (I) |
| $V_L$ | Resting potential | -70 mV |
| $V_{rest}$ | Reset potential | -55 mV |
| $V_{th}$ | Spike threshold | -50 mV |
| $V_E$ | Reversal potential of excitatory synapses | 0 mV |
| $V_I$ | Reversal potential of inhibitory synapses | -70 mV |
| $\tau_{ref}$ | Refractory period | 2 ms (E), 1 ms (I) |
| $\tau_E$ | Excitatory synaptic time constants | 5 ms |
| $\tau_I$ | Inhibitory synaptic time constants | 10 ms |
| $\tau_d$ | Delay of synaptic transmission | 1 ms |
| $I_0$ | Bias current | 0.5 nA (E), 0.3 nA (I) |
| $\sigma$ | Fluctuation intensity of the background current | 0.1 nA |

**Figure legends**

**Figure 1: Schematic presentation of the neural circuit model.** A: The neuronal network comprises $N_E$ excitatory and $N_I$ inhibitory neurons. Neurons in the network are coupled via excitatory (blue) and inhibitory (red) synapses governed by different coupling strengths. Each neuron also receives an external stimulus to maintain the network activity. B: Either the input heterogeneity (top) or output heterogeneity (bottom) is introduced into the network. In simulations, these two types of structural heterogeneity are controlled by the parameters $(K_{in}, S_{in})$ and $(K_{out}, S_{out})$, respectively. C: As examples, three typical in-degree distributions (Exponential, Gaussian and Uniform) are plotted, with the mean in-degree $K_{in} = 400$. A large value of $S_{in}$ indicates a strong level of structural heterogeneity.

**Figure 2: Multi-scale dynamics of neuronal networks at different levels of structural heterogeneity.** A: Dynamical performance for a network with heterogeneous input connectivity. Top panel: The input heterogeneity level over a simulated time of 1500 ms. Middle panel: Spike raster for a sample of 400 excitatory neurons (blue) and 100 inhibitory neurons (red). Bottom panel: Synaptic currents of a randomly chosen excitatory neuron. Here different colors represent different types of synaptic currents: excitatory (blue), inhibitory (red), and the total of excitatory and inhibitory (black). The dark shaded region surrounding the black curve denotes the error bounds (standard deviation) of the total excitatory and inhibitory current over all excitatory neurons. The collective firing of neurons exhibits distinct population synchrony degrees at different levels of input heterogeneity. The light shaded regions in (A) denote the state transitions. B: The corresponding dynamical performance for a network with heterogeneous output connectivity is plotted, in which the collective neuronal firing displays high synchrony at different levels of output heterogeneity. In simulations, three different levels of structural heterogeneity are considered: $S_{in}$ (or $S_{out} = 0$), $S_{in}$ (or $S_{out} = 220$), and $S_{in}$ (or $S_{out} = 380$), respectively.

**Figure 3: Dynamical performance of neuronal networks at different levels of structural heterogeneity.** Typical neuronal measurements are plotted as a function of the level of structural heterogeneity. Here four measurements are considered: the average standard deviation of the total

synaptic currents per time instant (A), the coherence index (B), the mean $CV_{ISI}$ (C) and the average firing rate (D). All data are plotted as the mean $\pm$ SD (standard deviation). In (A)-(D), the red color represents the network with input heterogeneity ($S_{in}$) and the blue color denotes the network with output heterogeneity ($S_{out}$).

**Figure 4: Identification of neuronal avalanches evoked by heterogeneity in synaptic inputs.**
A: Mapping spikes from $N_S$ randomly sampled excitatory neurons into time bins ($\Delta t$=0.5 ms). Here an avalanche event is defined as a sequence of time bins in which at least one spike is emitted, ending with a silent time bin. As an example, the $i$-th avalanche event with a size of 40 spikes and duration of 7.5 ms is illustrated in (A). B: Typical distributions of avalanche size (top) and avalanche duration (bottom) for networks with heterogeneous input connectivity ($N_S = 400$). At different levels of input heterogeneity, the model may present subcritical (red, $S_{in} = 0$), critical (black, $S_{in} = 240$), and supercritical (blue, $S_{in} = 380$) avalanche dynamics. Insert: at the intermediate level of input heterogeneity ($S_{in} = 240$), the average size $\langle S \rangle(T)$ conditioned on a given duration $T$ shows power-law increases corresponding to $\langle S \rangle(T) \sim T^{\pi}$. C: The corresponding distributions of avalanche events for networks with heterogeneous output connectivity. Similarly, three levels of output heterogeneity are considered: $S_{out} = 0$ (red), $S_{out} = 240$ (black), $S_{out} = 380$ (blue). The model exhibits only supercritical avalanche dynamics due to high synchrony among neurons.

**Figure 5: Neuronal avalanches are regulated by input heterogeneity.** A: Distance between real and best-fitted power-law distributions for avalanche size (blue) and duration (red) versus the level of input heterogeneity. The model exhibits subcritical (left region), critical (middle shaded region), and supercritical (right region) avalanche dynamics for different values of $S_{in}$. B: Best-fitted slopes for power-law distributions of avalanche size ($\alpha$, blue) and duration ($\beta$, red) versus the level of input heterogeneity. C: Distances in (A) are replotted as a function of the coherence index. D: Examples of the population firing rate and time-frequency wavelet spectrogram (left) and the average power spectrogram over time (right). The two levels of input heterogeneity considered here are $S_{in} = 200$ (top) and $S_{in} = 240$ (bottom). E: The peak frequency (top) and peak power (bottom) are computed at different levels of input heterogeneity.

**Figure 6: Effects of sampling size on the observation of neuronal avalanches.** A: Recalculated distributions of avalanche size at different sampling sizes. B: Corresponding distributions of avalanche duration at different sampling sizes. In (A) and (B), we use the spike data generated in the same experiment (Fig. 4b, $S_{in} = 240$). The five sampling size considered here are $N_S = 200, 300, 400, 600,$ and $800$. C: Distance between real and best-fitted power-law distributions for avalanche size (blue) and duration (red) as a function of sampling size. The model exhibits the critical (shaded region) avalanche dynamics for an intermediate sampling size. D: Best-fitted slopes for power-law distributions of avalanche size ($\alpha$, blue) and duration ($\beta$, red) versus sampling size.

**Figure 7: Neuronal avalanches can be evoked in networks with both Gaussian and uniform in-degree distributions.** A: Typical distributions of avalanche size (left) and avalanche duration (right) for networks with Gaussian in-degree distributions. At different levels of input heterogeneity, the model displays subcritical (red, $S_{in} = 0$), critical (black, $S_{in} = 200$), and supercritical (blue, $S_{in} = 380$) avalanche dynamics. Insert: at an intermediate level of input heterogeneity ($S_{in} = 200$), the average size $\langle S \rangle (T)$ conditioned on a given duration $T$ shows power-law increases corresponding to $\langle S \rangle (T) \sim T^{\pi}$. B: The corresponding distributions of avalanche events for networks with uniform in-degree distributions. Similarly, the model exhibits different types of dynamical behaviors at different levels of input heterogeneity. In simulations, we set the sampling size at $N_S = 400$.

**Figure 8: Regulations of neuronal avalanches by input heterogeneity in networks with Gaussian and uniform in-degree distributions.** A: For the Gaussian in-degree distributions, we plot the distance measurements between real and best-fitted power-law distributions (left) and the best-fitted slopes (middle) as a function of the level of input heterogeneity. In addition, the distance measurements for avalanche size and duration are replotted versus the coherence index

(right). In (A), the blue lines denote the measurements for avalanche size, and the red lines represent the measurements for avalanche duration. B: The corresponding plots for networks with uniform in-degree distribution. In comparison with the model with uniform in-degree distribution, it is obvious that the model with the Gaussian in-degree distribution displays a relatively wider and right-shifted criticality region.

**Figure 9: Neuronal oscillations are modulated by structural heterogeneity.** A: For networks with Gaussian in- and out-degree distributions, both the dominant (peak) frequency (left) and the corresponding peak power (right) are plotted as a function of the level of structural heterogeneity. B: The corresponding results for networks with uniform in- and out-degree distributions. The red color represents the network with input heterogeneity ($S_{in}$) and the blue color denotes the network with output heterogeneity ($S_{out}$).

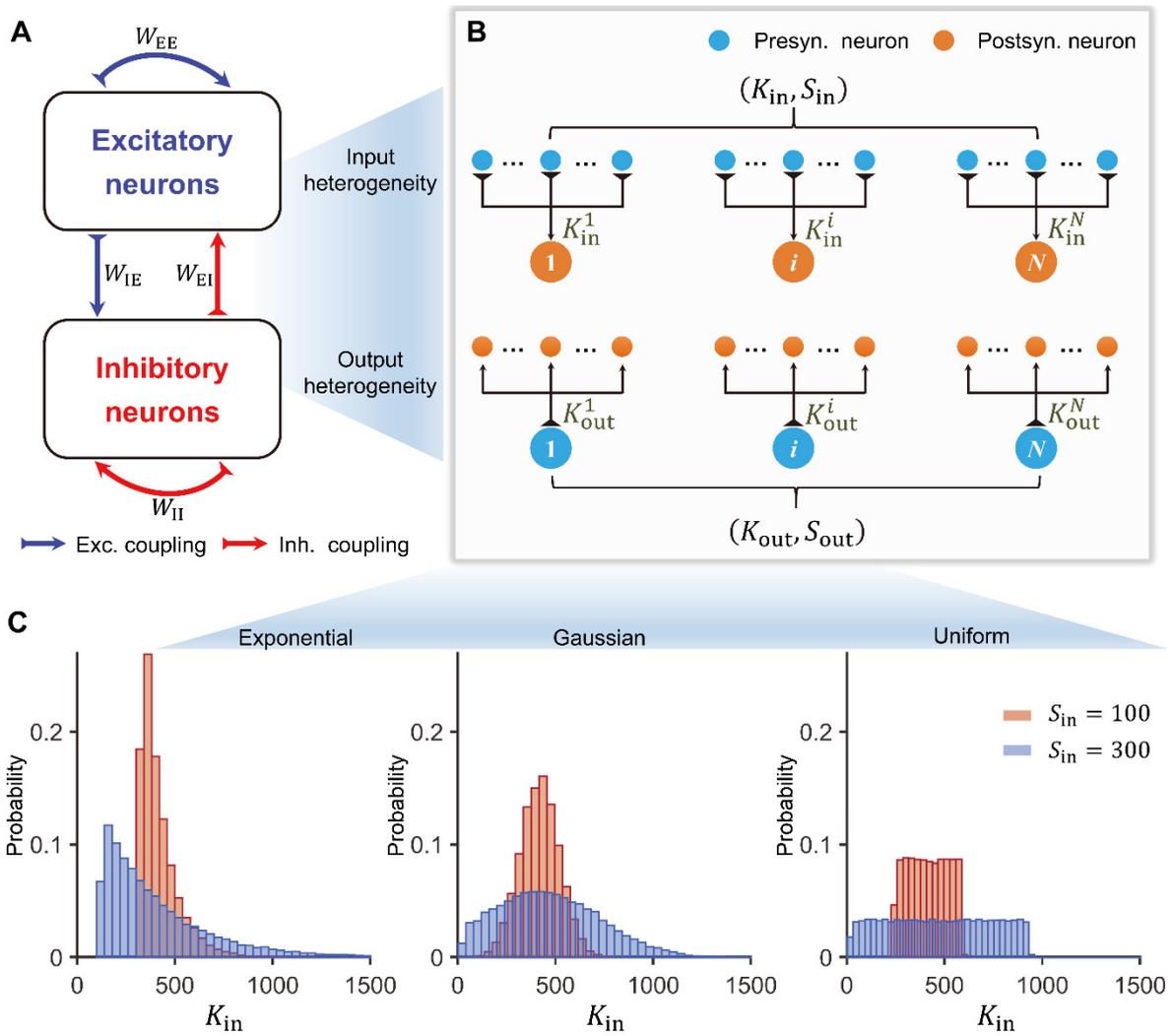

**Figure 1**

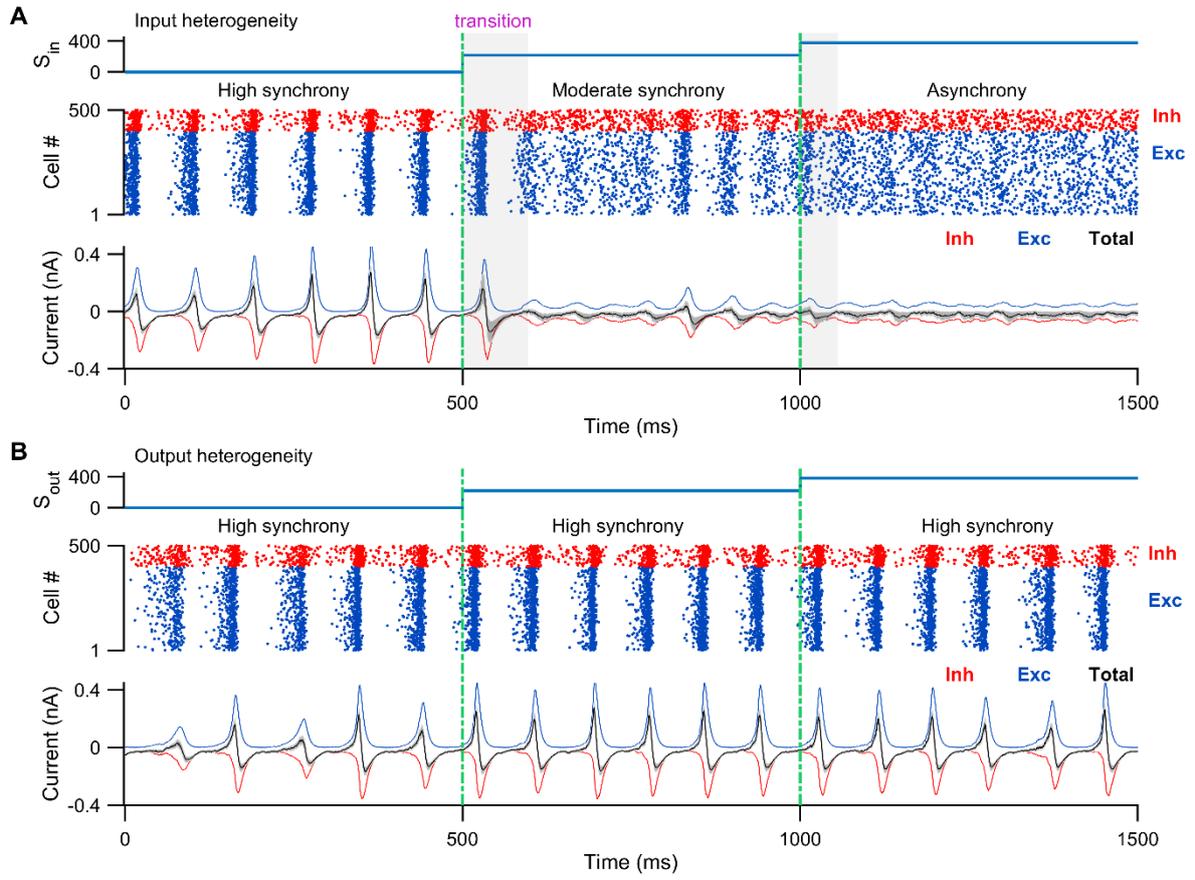

**Figure 2**

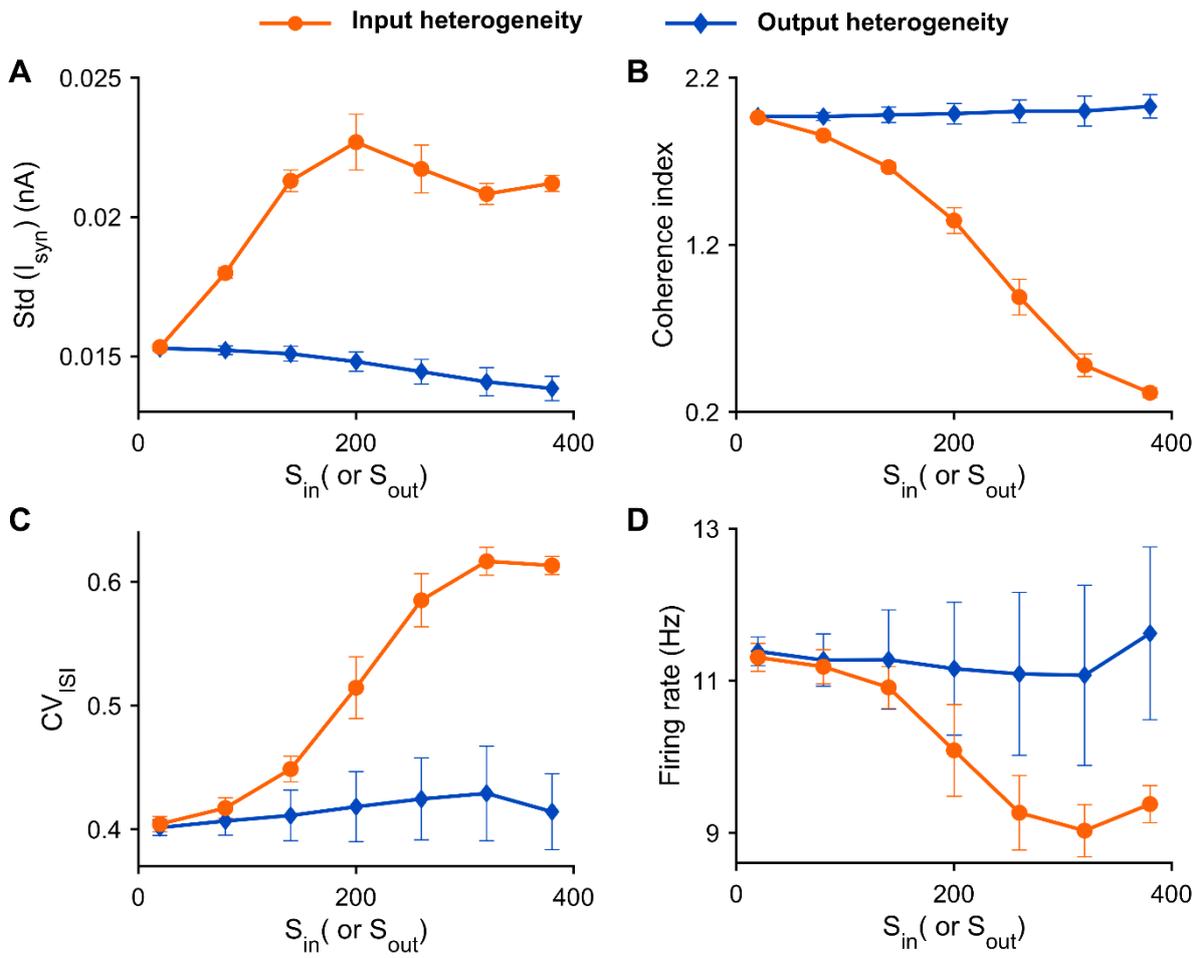

**Figure 3**

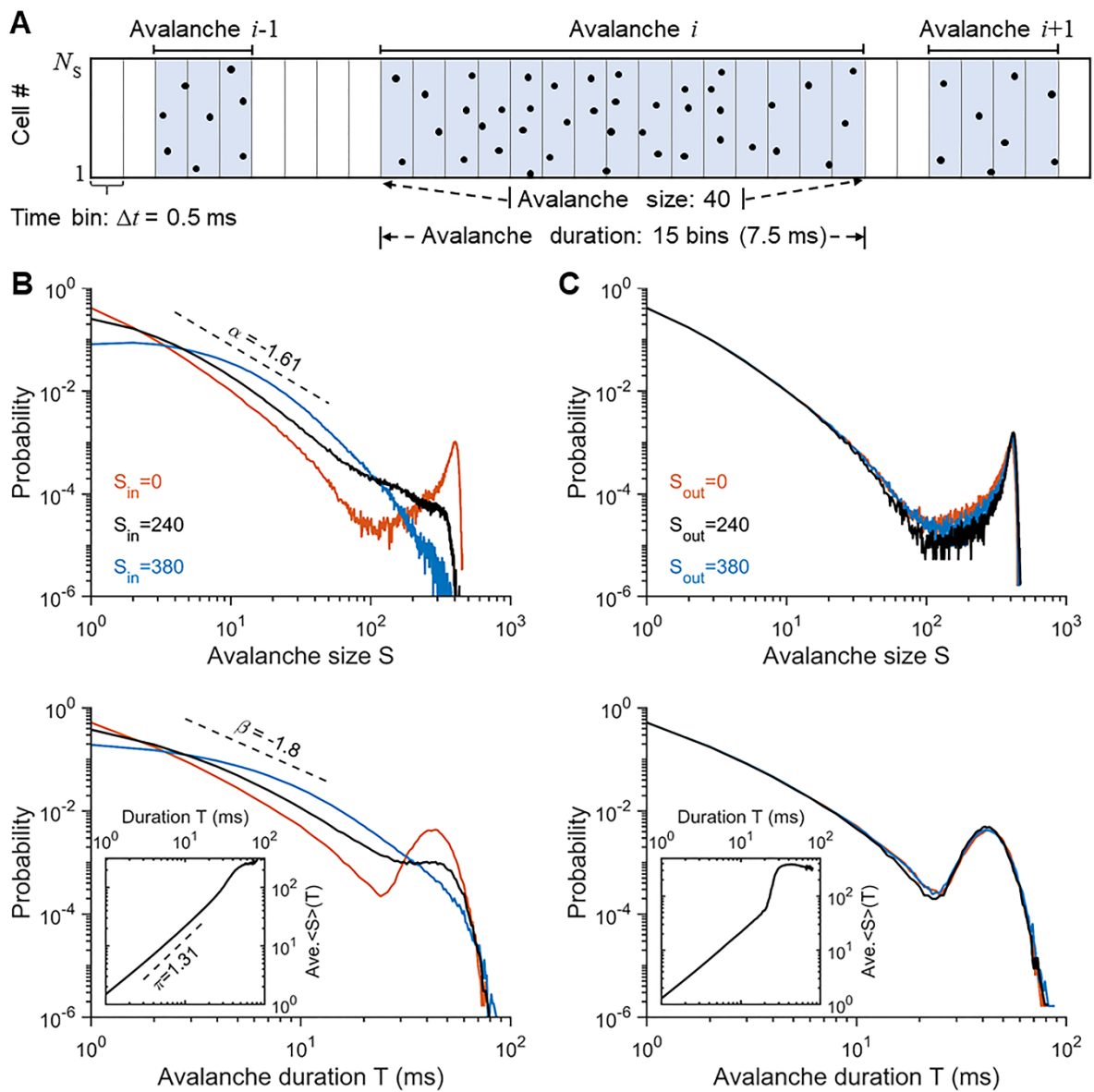

**Figure 4**

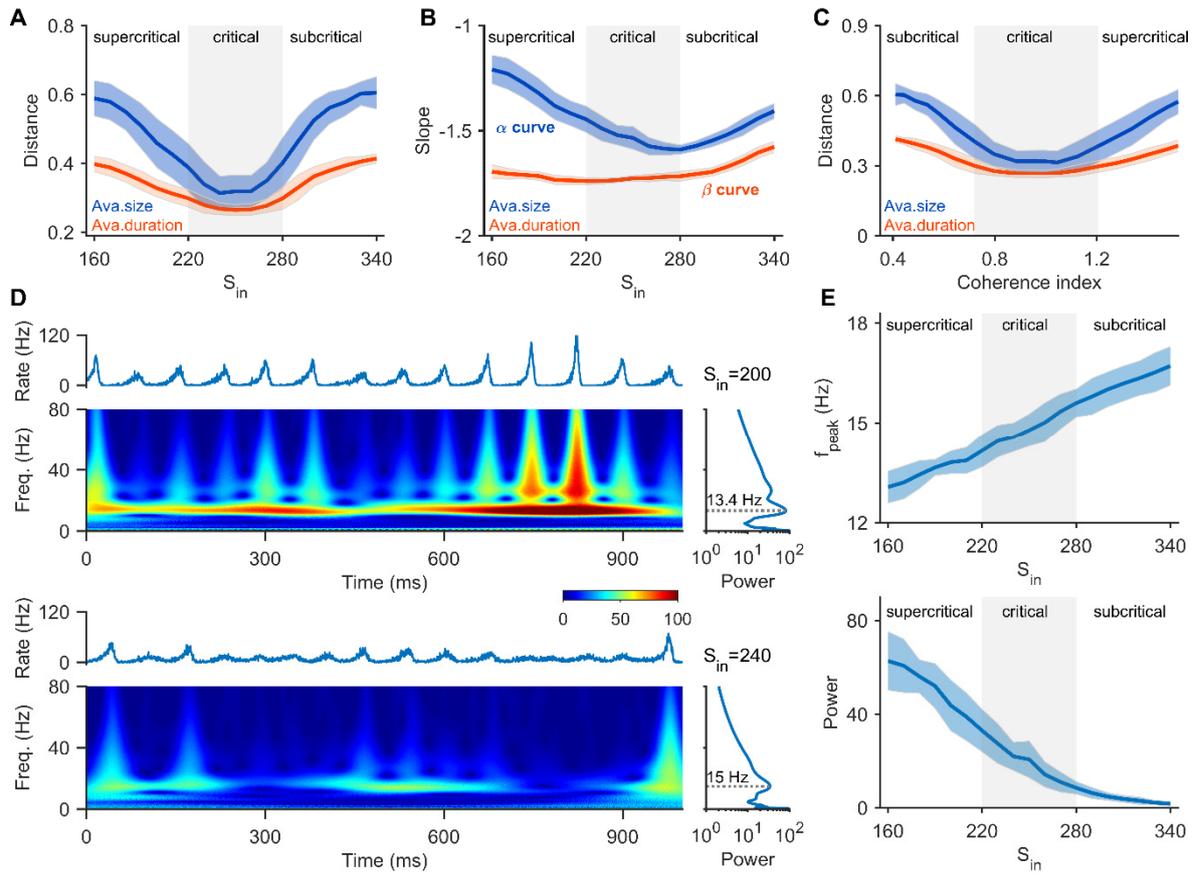

**Figure 5**

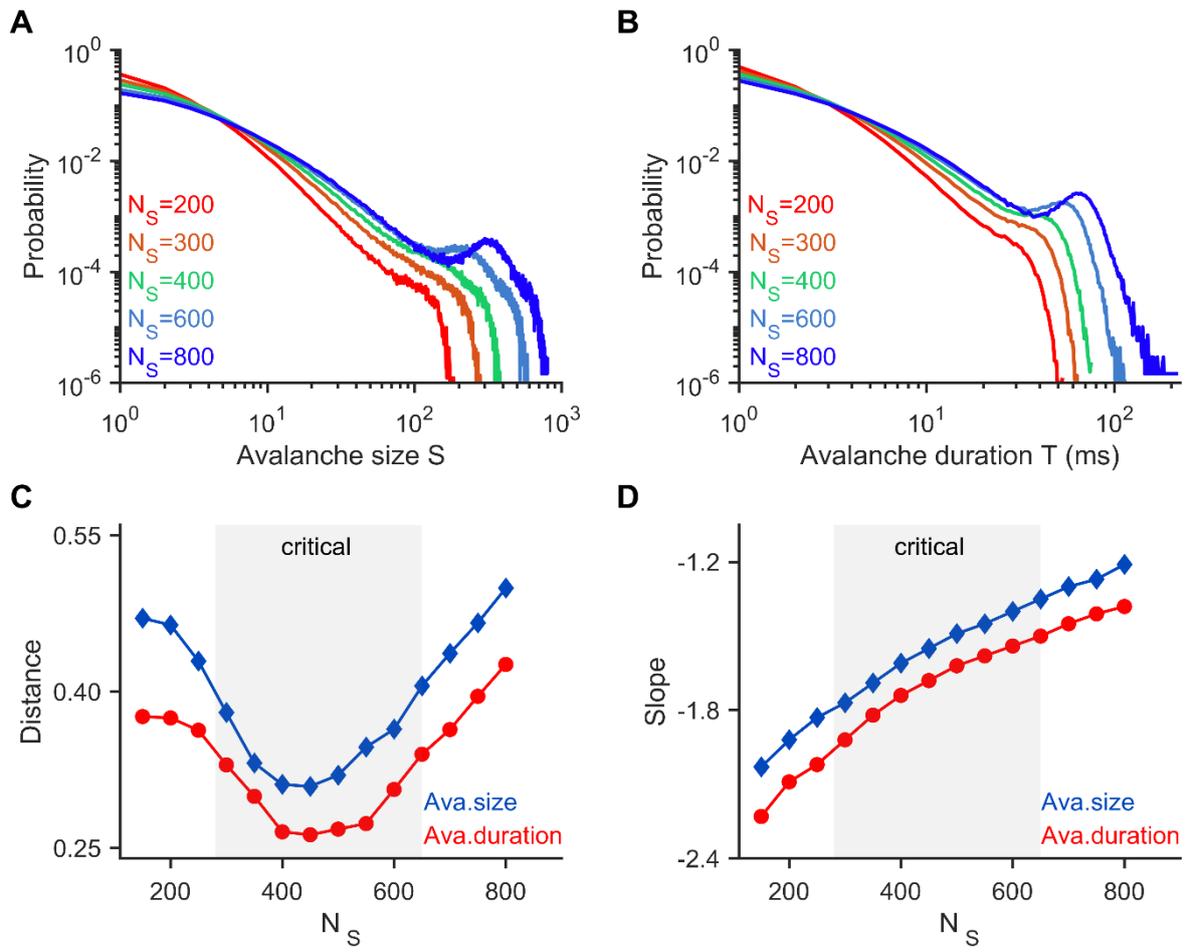

**Figure 6**

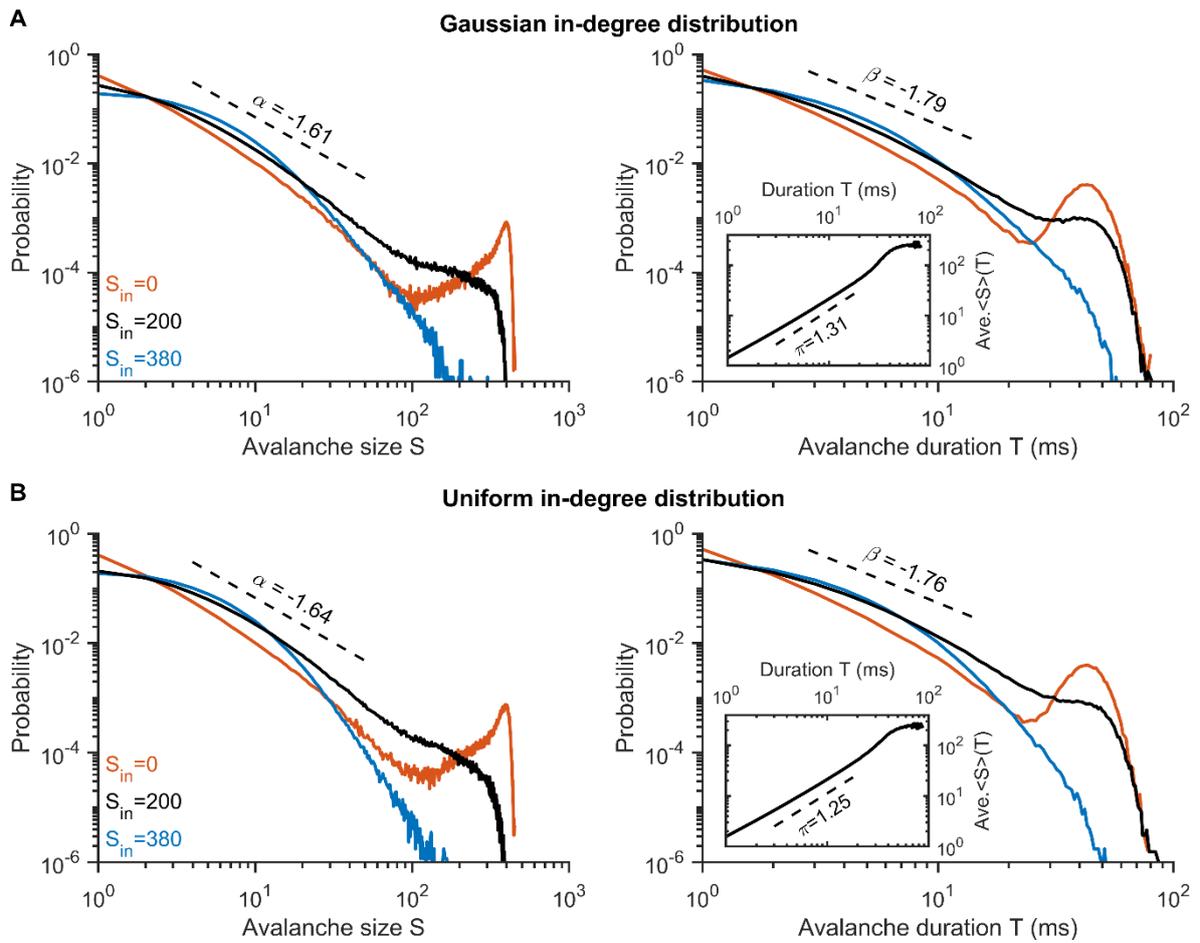

**Figure 7**

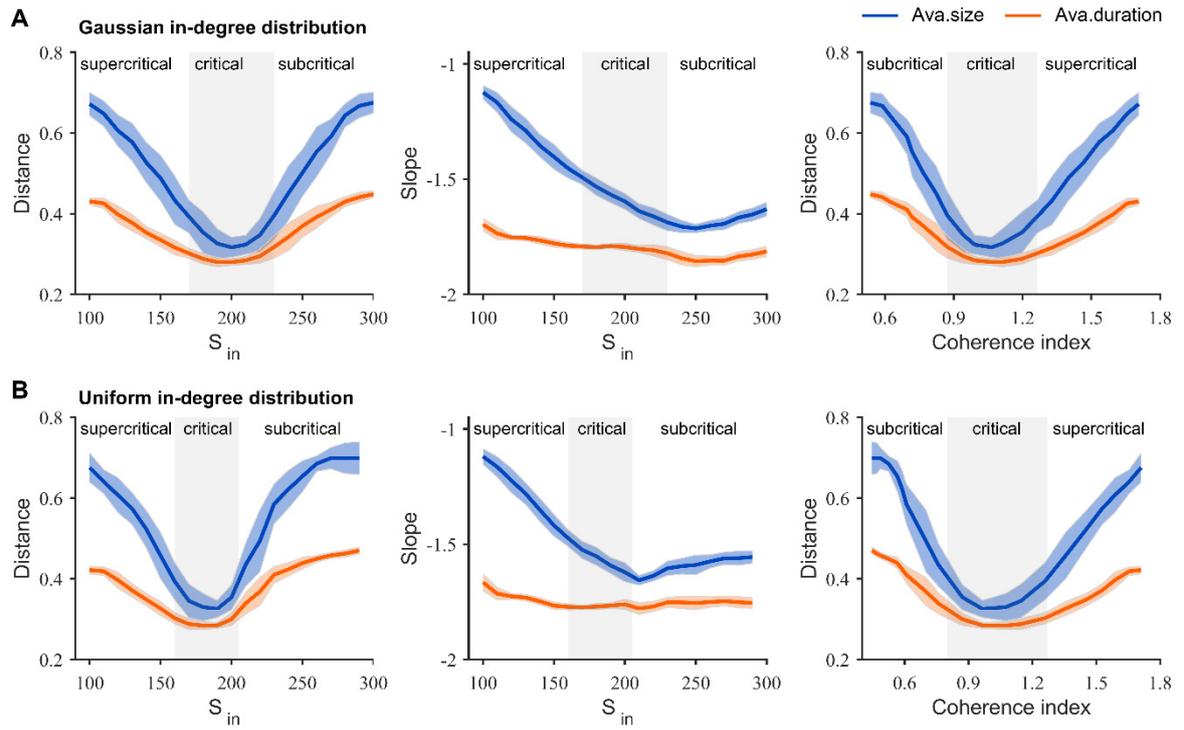

**Figure 8**

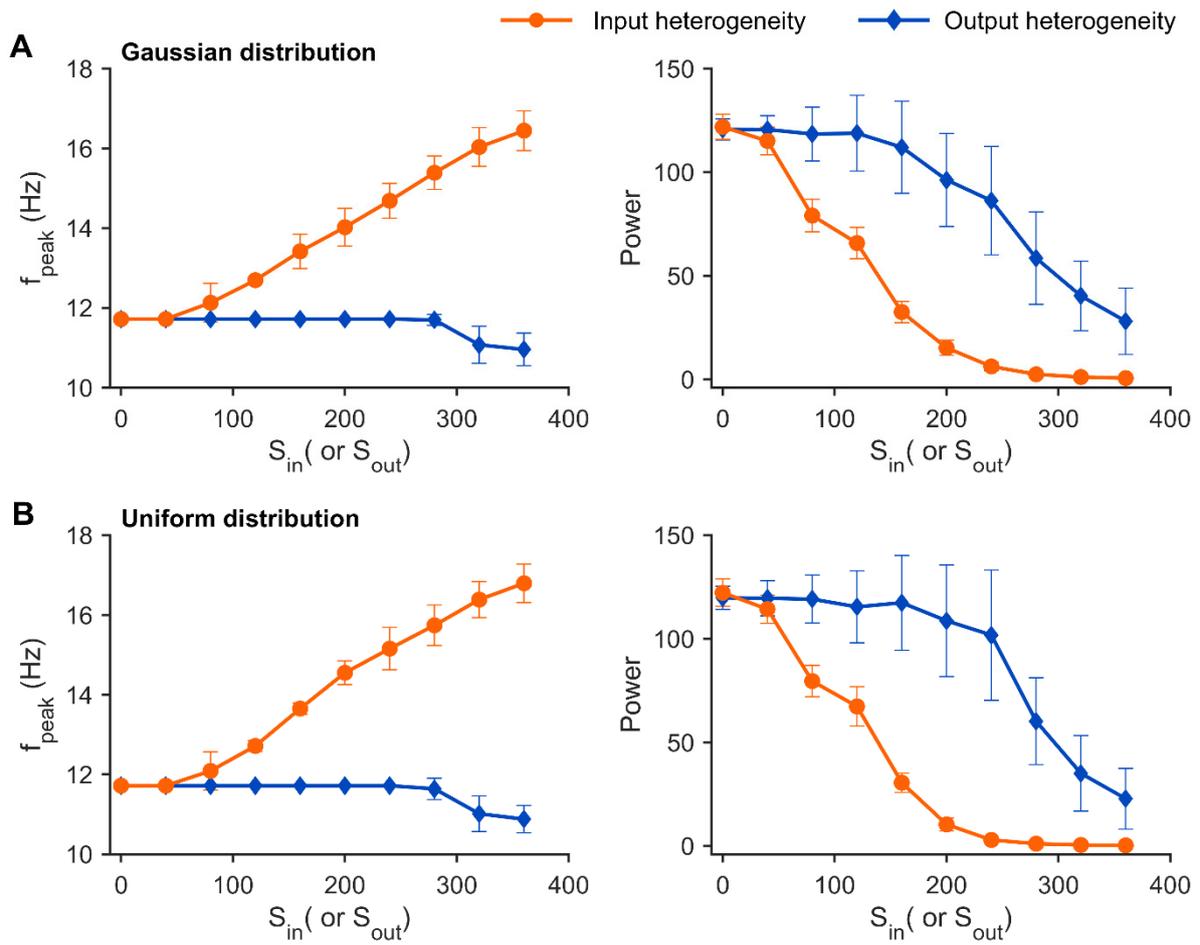

**Figure 9**